# A comprehensive study of distribution laws for the fragments of Košice meteorite

Maria Gritsevich [1, 2, 3], Vladimir Vinnikov [2], Tomáš Kohout [4, 5], Juraj Tóth [6], Jouni Peltoniemi [4,1], Leonid Turchak [2], Jenni Virtanen [1]

1. *Finnish Geodetic Institute, Geodeetinrinne 2, P.O. Box 15, FI-02431 Masala, Finland*
2. *Russian Academy of Sciences, Dorodnicyn Computing Centre, Department of Computational Physics, Vavilova ul. 40, 119333 Moscow, Russia*
3. *Lomonosov Moscow State University, Institute of Mechanics, Michurinsky prt., 1, 119192, Moscow, Russia*
4. *University of Helsinki, Department of Physics, P.O. Box 64, 00014 Helsinki University, Finland*
5. *Academy of Sciences of the Czech Republic, Institute of Geology, Rozvojová 269, 16500 Prague 6, Czech Republic*
6. *Comenius University, Faculty of Mathematics, Physics and Informatics, Mlynská dolina, SK-84248 Bratislava, Slovakia*

## Abstract

In this study, we conduct a detailed analysis of the Košice meteorite fall (February 28, 2010), in order to derive a reliable law describing the mass distribution among the recovered fragments. In total, 218 fragments of the Košice meteorite, with a total mass of 11.285 kg, were analyzed. Bimodal Weibull, bimodal Grady and bimodal lognormal distributions are found to be the most appropriate for describing the Košice fragmentation process. Based on the assumption of bimodal lognormal, bimodal Grady, bimodal sequential and bimodal Weibull fragmentation distributions, we suggest that, prior to further extensive fragmentation in the lower atmosphere, the Košice meteoroid was initially represented by two independent pieces with cumulative residual masses of approximately 2 kg and 9 kg respectively. The smaller piece produced about 2 kg of multiple lightweight meteorite fragments with the mean around 12 g. The larger one resulted in 9 g of meteorite fragments, recovered on the ground, including the two heaviest pieces of 2.374 kg and 2.167 kg with the mean around 140 g. Based on our investigations, we conclude that two to three larger fragments of 500-1000g each should exist, but were either not recovered or not reported by illegal meteorite hunters.

## 1. Introduction

The Košice fireball appeared in the night sky over the Central Europe on February 28[th], 2010 at 22:24:46 UT. As is usual for such notable cases (see e.g. the case of Přibram meteorite described in details by Ceplecha, 1961), a glare of the bolide illuminated ground at some places in Eastern Slovakia. Cannon-like burst or series of low frequency blasts were also reported (Tóth et al., 2014). Due to the cloudy skies, there were no direct fireball observations made by the European Fireball Network or the Slovak Video Network. Fortunately several surveillance cameras from Hungary captured the fireball (Borovička et al., 2013). Trajectory analysis of these records, performed by Jiri Borovička, lead to the conclusion, that significant number of fragments has survived the atmospheric entry and reached the ground forming strewn field near the city Košice in Eastern Slovakia. The data from Local Seismic Network Eastern Slovakia confirmed the derived Košice fireball atmospheric trajectory. The following search campaign was successful. The first meteorite fragment was discovered by Juraj Tóth on March 20th within the predicted area. In total 78 meteorite fragments were found in expeditions organized by the Comenius University in Bratislava (under the leadership of Juraj Tóth), Astronomical Institute of the Slovak Academy of Sciences (under the leadership of Jan Svoren), and Czech Academy of Sciences (under the leadership of Pavel Spurný). Also taking into account also meteorite fragments that were found during illegal searches, total 218 fragments of the Košice meteorite were recognized (Appendix I, Table 1). This makes Košice one of the richest instrumentally-recorded meteorite falls to date in terms of number of recovered fragments (Appendix I, Table 2). The two heaviest fragments weigh 2.374 kg and 2.167 kg. The current total known mass of the Košice meteorite is 11.285 kg with almost 7 kg belonging to the collection of the Comenius University in Bratislava and Astronomical Institute of Slovak Academy of Sciences. The meteorites masses and positions collected in the strewn field were reported to Juraj Tóth by different searchers. After their evaluation, they were included in the list of meteorites. Laboratory analysis revealed that Košice meteorite is an ordinary H5 chondrite with average bulk and grain density 3.43 g/cm$^3$ and 3.79 g/cm$^3$ respectively which have resulted from the homogenous parent body (Kohout et al., 2014).

Hypervelocity atmospheric entry is a complex multistage physical process lacking straightforward mathematical description due to the large number of factors. Ground-based meteor observations do not directly provide us with all necessary information needed for modeling, such as, for example, meteoroid bulk and grain densities, number and shape of individual meteoroid fragments and their homogeneity. Thus, numerical simulations are often computationally extensive and cover only few possible scenarios. Examples are an assumption that each fragmentation event of the meteoroid spawns a number of almost equal pieces (Borovička and Kalenda, 2003), or consideration of constant main body shape along the whole luminous segment of its trajectory (Halliday et al., 1996). More recent studies are carried out by direct numerical computations with account for underlying physics with larger number of free parameters and/or key parameters varying along the trajectory (Ceplecha and ReVelle, 2005; Gritsevich et al., 2011; Dergham, 2013). Practical implementations usually demand simplification of the used mathematical models, though analytical approach helps to avoid running into the uniqueness problem by keeping simple set of free independent parameters (Gritsevich, 2008; Bouquet, 2013).

Several numerical models focused on meteoroid's fragmentation were implemented. Ivanov and Ryzhanskii (1999) modeled the meteoroid breakup as sequential dichotomy of the parent body. Artemieva and Shuvalov (1996) considered aerodynamic interaction between equally sized fragments. In later study, Artemieva and Shuvalov (2001) pointed out the stochastic nature of cracks in the meteoroid body, obeying the statistical strength theory and utilized the Weibull distribution to divide a meteor body into the non-equal pieces.

The fragment mass distribution for the Košice meteorite was briefly discussed by Borovička et al. (2013). Based on the observed light curve and deceleration of the Košice bolide the authors estimated the meteoroid pre-atmospheric mass as 3500 kg. The authors also describe in great details possible scenario of its fragmentation in the atmosphere. According to the model, the first significant fragmentation occurred most likely in two phases at heights of 57–55 km. The authors question the number of recovered meteorite fragments in the range 100 g - 1000 g, exceeding by almost 50% their model prediction (17 recovered meteorites against 12 according to their model). The modeled and actual strewn fields are also provided (Borovička et al. 2013, Fig. 18). It is suggested that the outlying fragments may be formed at higher altitudes and thus were not considered in the applied model. The authors note, that the number and masses of small fragments are not well retrievable from the light curve analysis, since different combinations of fragment mass range and their mass distribution could be used to fit the flares.

In the following sections, we mathematically test popular distribution laws on the available Košice meteorite fragment data. We start with a normal and lognormal distribution of fragment masses in logarithmic and linear scales, respectively. Next, we consider more advanced distributions commonly used in fragmentation theory. These functions with carefully calibrated parameters were successfully applied earlier (Elek and Jaramaz, 2009, Silvestrov, 2004ab). As a result we obtain integral characteristics of the underlying fragmentation event. At the end we make predictions of meteoroid mass distributions following fragmentation events. These recommendations can be used, in particular, to generate most realistic population of fragments in the dark flight numerical simulations for meteorite-producing fireballs.

## 2. Application of selected statistical models

### 2.1. Methodology and simple distributions

The masses of the Košice meteorite fragments analyzed in this study are listed in the Appendix I, Table 1. Since fragment masses span over four orders of magnitude, it is convenient to operate with them in the logarithmic scale introducing non-dimensional mass values, further referred to as $m$, by scaling all masses given in Table 1 to one gram (e.g. for the largest fragment with mass 2374 g we obtain $\ln(2374) \approx 7,77$). Figure 1 shows empirical cumulative distribution function (CDF) for this data set. The experimental points on the plot have clearly distinguishable shape, which can be fitted by a number of distributions, including normal, logistical and other continuous sigmoid cumulative functions as Weibull distribution. For the sake of brevity and simplicity we consider only distributions commonly used to describe and simulate of various fragmentation processes such as grinding, milling and impact crushing of hard materials (Elek and Jaramaz, 2009, Silvestrov, 2004ab). Usually these functions provide satisfactory approximation and it is highly unlikely to reproduce fragmentation sequences or even compose similar shaped sigmoid CDFs using more rare distributions.

For example, the observed distribution can be approximated via normal CDF for the number of fragments, which has the form:

$$F\left(x, \mu, \sigma^2\right) = \int_{-\infty}^{x} \frac{1}{\sqrt{2\pi\sigma^2}} e^{-\frac{(t-\mu)^2}{2\sigma^2}} \, dt \, .$$

The mean $\mu$ and the unbiased standard deviation $\sigma$ are a-priori unknown and computed from the logarithmically transformed sample as:

$$\mu = \frac{1}{N} \sum_{i=1}^{N} \ln(m_i), \ \sigma = \sqrt{\frac{1}{N-1} \sum_{i=1}^{N} \left[\ln(m_i) - \mu\right]^2} \ , \text{ where N – is the sample size.}$$

In order to confirm or reject goodness of fit test of various selected theoretical distributions, we use Pearson's chi-squared test. The main idea behind this test is to compute a normalized relative error between the obtained sample and assumed distribution to compare it against chi-square quantile with desirable significance level. The test is described in details in the Appendix II.

As a rule, two requirements should be met to conduct the Pearson's test properly. The first one requires that the sample size $N$ should be large enough ($N \geq 30$) and the second one assumes that the expected frequencies $N_k$ should not be less than five (Turin and Makarov, 1998).

It is important to keep in mind that the null-hypothesis about assumed distribution is not a simple, but a composite one, since CDF parameters (e.g. mean and standard deviation) are a-priori unknown. The above mentioned estimations of these parameters are not applicable directly to the chi-squared test. Instead, techniques of maximum likelihood is applied. One can first take a sample mean and unbiased sample standard deviation as initial values for the required parameters, then vary them to "minimize" the chi-squared distance $\chi^2_{emp}$ (as stated in Fisher's theorem), and then compare obtained "minimal" distance to the desired chi-squared quantile.

We consider following parameters for the Pearson's test on the sample studied. The close ends of the interval $L$ are taken as $x_0 = \ln(m_1) - \varepsilon$, $x_K = \ln(m_N) + \varepsilon$, where $\varepsilon$ is a small number. Each subinterval $L_k$ is constructed in such a way that its estimated sample frequency $\nu_k$ is no less than

10. For the chi-squared quantile we take $\alpha = 0.05$. Normal distribution has two independent parameters, so $p = 2$. Our set of data (further referred to as observed distribution) yields the results summarized in the Appendix III, table 1.

The result for empirical chi-squared statistics is: $\chi^2_{emp} = 14.68$ and chi-squared quantile equals: $\left(\chi^2\right)^{-1}(\alpha, K - p - 1) = 26.3$. Thus, the null-hypothesis for normal distribution can be accepted since it is valid with 95% probability. Taking into account that a continuous probability distribution of a random variable, whose logarithm is normally distributed, is a lognormal distribution, our finding looks quite reasonable and is in agreement with (Kolmogorov, 1941).

However, the first obtained sample permits other similar-shaped continuous distributions, for example logistic function:

$$F(x, \mu, s) = \frac{1}{2} + \frac{1}{2}\tanh\left(\frac{x - \mu}{2s}\right), \text{ where } s = \sigma\sqrt{3}/\pi.$$

In this case, Pearson's chi-squared statistics is even less and equals: $\chi^2_{emp} = 14.26$. The logistic CDF often considered as a more simple form not involving integration for normal distribution.

There are several other goodness-of-fit tests. For the normal distribution we also use modified Kolmogorov–Smirnov test (Turin and Makarov, 1998) and Wald–Wolfowitz runs test (Hauck, 1971) (see Appendix II).

For the present sample and the above mentioned sample mean $\mu = 2.52$ and unbiased standard deviation $\sigma = 1.41$ we obtain $\left(\sqrt{N} - 1 + \frac{0.85}{\sqrt{N}}\right)D = 0.664$. The value of the Stephens corrected Kolmogorov quantile at level $\alpha = 0.05$ is $0.895$. This indicates that null-hypothesis is also accepted under Kolmogorov–Smirnov test. The use of other tests is highly dependent on their complexity. Among simple tests one can also apply G-test (Appendix II), which is slightly more accurate than Pearson chi-squared statistics.

Regardless of the null-hypothesis acceptance for lognormal distribution, we consider other CDFs for better goodness-of-fit values. First, a histogram is constructed from the recovered masses. The histogram shape reveals that the sample points satisfy superposition of two or more distributions.

The plot on Fig. 2 shows no significant secondary peaks, though the sample exhibits small local maxima on the left and right tails. In our case this form of statistical representation offers little insight into the possible multimodal nature of the underlying theoretical distribution. Still, there is another approach to statistical investigation of meteorite fragments. It deals with the complementary cumulative number of fragments $N$ instead of the normalized cumulative number of fragments $F(m)$ as in classical CDF.

First it is crucial to determine if collected data are statistically valid. To check this, we construct the $\log(N(\geq m))$ vs. $\log(m)$ plot, where $N(\geq m)$ is the number of fragments with mass greater than m (Fig.3). Since our model functions and experimental data are often quite similar, we also present the difference plots (Fig. 10). One can observe a significant gap of data for massive fragments. Actually, the two last points stand aside from remaining empirical distribution and correspond to the weak local maximum on the Fig.2. It is a matter for discussion whether to consider this maximum as a second mode, since two data points do not provide any feasible statistics. On contrary, these two data points do not distort goodness of fit test significantly. There are still few gaps in the mass samplings which can serve as possible delimiters for various modes.

The common practice for constructing multimodal distribution is to limit the number of modes. Usually bimodal and trimodal CDFs are chosen:

$F_{BM}(x) = (1-\omega)F_1(x) + \omega F_2(x)$,    $F_{TM}(x) = (1-\omega_2-\omega_3)F_1(x) + \omega_2 F_2(x) + \omega_3 F_3(x)$,    where    $\omega_\bullet$    –    are appropriate weight coefficients. $\omega \in [0;1]$, $\omega_2 \geq 0$, $\omega_3 \geq 0$, $\omega_2 + \omega_3 \in [0;1]$.

The reason for such limitation is following. The number of independent parameters for the distribution increases with each additional mode. While such flexibility can be handy to approximate given samples, it also decreases the total degrees of freedom for the chi-squared test and lowers the threshold of the quantile.

We apply bimodal lognormal function to the sample. Since such function has two means, two deviations and the weighting coefficient, the resulting shape can be tuned to fit sample data with better accuracy. Minimizing the functional of empirical chi-squared statistics we obtain suboptimal values for independent parameters $\omega = 0.9$ $\mu_1 = 2.22$, $\mu_2 = 5.19$ and the standard deviations $\sigma_1 = 1.05$, $\sigma_2 = 0.88$, that provide empirical chi-square estimation equal to $\chi^2_{emp} = 9.44$ which is well below the threshold of 23.36 for the number of independent parameters $p = 5$, and the number of subintervals $K = 19$. This distribution better follows original sample points than unimodal one (see Fig. 4-5). The effective minimization procedure is a complex problem especially for multidimensional multimodal case. For the described here problem, it is sufficient to proceed with manual iterations by introducing sample mean and sample variance (e.g. taking already estimated parameters from other distribution) and alternating the descent steps for mean and variance. This kind of search is appropriate for the sigmoid functions. The goal of the procedure is to reduce the chi-square estimation $\chi^2_{emp}$ below the five percent quantile to confirm the underlying null-hypothesis. We leave questions of uniqueness of solution and standard deviation of sought parameters outside of the scope of the present paper. Alternative techniques can be found e.g. in (Elek and Jaramaz, 2009).

## 2.2.     Advanced statistical models

Various aspects of fragmentation processes are quite commonly discussed in scientific literature (e.g. Kolmogorov, 1941, Gilvarry, 1961, Grady, 1985). To simplify our approach we omit theoretical issues involving Rosen – Rammler equation and its implications and focus on mainstream statistical distributions. Lesser known distributions, such as generalized Mott and Held distributions, are discussed e.g. in the paper by Elek and Jaramaz (2009).

Apart from above-mentioned extensively used lognormal function, there are other well-known statistical laws dealing with fragmentation and size distribution of particles.

The Weibull distribution provides successful empirical description for lifetimes of objects, fatigue data and the size of particles generated by grinding, milling and crushing operations. Moreover, Martin et al. (1980) showed that Weibull function describes a mass distribution for chondrules disaggregated from the various meteorites with considerable precision. The CDF for the mass of fragments has the form:

$F_W(m, \gamma, \mu) = \dfrac{M(\leq m)}{M_0} = 1 - \dfrac{M(\geq m)}{M_0} = 1 - \exp\left(-\left(\dfrac{m}{\mu}\right)^\gamma\right)$,    where $M_0$ is the total mass of the obtained distribution.

The linear exponential distribution, known as Grady distribution (Grady and Kipp, 1985; Grady, 1990) represents cumulative number of fragments as:

$$F_{GK}(m, \mu) = 1 - \frac{N(\geq m)}{N_0} = 1 - \exp\left(-\frac{m}{\mu}\right), \text{ where } N_0 = \frac{M_0}{\mu}.$$

The Gilvarry distribution (Gilvarry, 1967, Gilvarry and Bergstrom., 1967) is proposed for the same purpose as before-mentioned statistical functions. However, it is given as the dimensional probability density. For example, its one dimensional form is:

$$f_G(m) = \frac{1}{\mu}\frac{M_0}{m}\exp\left(-\frac{m}{\mu}\right)$$

According to laboratory experiments, Gilvarry distribution describes breaking with fine fragments excessively contributing to the size spectrum, which, practically, disperse in the atmosphere and cannot be recovered. Cumulative distribution can be obtained by integrating $f_G(m)$:

$$F_G(m) = \int_0^m \frac{1}{\mu}\frac{M_0}{x}\exp\left(-\frac{x}{\mu}\right)dx$$

This integral is diverging since the exponential integral $Ei(x)$ has a singularity at zero argument (see e.g., Gritsevich and Koschny, 2011). However this divergence occurs only as a mathematical formalism. Indeed the number of fragments $f_G(m)$ goes to infinity as their individual masses $m$ go to zero. But the total mass remains bounded:

$$M_G(m) = \int_0^m \frac{M_0}{\mu}\exp\left(-\frac{x}{\mu}\right)dx = M_0 \cdot \left(1 - \exp\left(-\frac{m}{\mu}\right)\right).$$

In reality the fragmentation is discrete, so there is constraint on a minimal fragment mass. If we introduce such constraint $m_0$ to the $f_G(m)$, we get

$$f_G(m) = \begin{cases} \frac{1}{\mu}\frac{M_0}{m}\exp\left(-\frac{m}{\mu}\right), & m \geq m_0, \\ 0, & m < m_0 \end{cases}$$

This leads to convergence of complementary CDF:

$$F_G(m) = \int_{m_0}^m \frac{1}{\mu}\frac{M_0}{x}\exp\left(-\frac{x}{\mu}\right)dx$$

This integral can be also expanded into the series as:

$$F_G(m) = \frac{M_0}{\mu}\left[\ln(x) + \sum_{k=1}^{\infty}\frac{(-x/\mu)^k}{k \cdot k!}\right]_{m_0}^m,$$

It is more convenient to integrate probability density function numerically by trapeze method on sufficiently fine grids. The setback of this distribution is the arbitrariness of cut-off mass $m_0$. It is prudent to take it as a minimal single mass of recovered meteorite fragments, i.e. 0,3 g for the Košice meteorite. One can also apply this limit with account for the smaller unrecovered particles.

These above mentioned distributions in their bimodal forms are applied to the obtained data. The resulting values of the Pearson's chi-squared for each of respective distributions are summarized in the Table 1. The values of Kolmogorov-Smirnov test and Wald–Wolfowitz runs test are given in Appendix 2.

First, we investigate the bimodal Weibull distribution:

$$F_W\left(m, \omega, \gamma_1, \mu_1, \gamma_2, \mu_2\right) = \omega\left[1 - \exp\left(-\left(\frac{m}{\mu_1}\right)^{\gamma_1}\right)\right] + (1-\omega)\left[1 - \exp\left(-\left(\frac{m}{\mu_2}\right)^{\gamma_2}\right)\right].$$

The parameters for this CDF are the weighting factor $\omega$, the shape $\gamma_1$ and scale $\mu_1$ for the first mode and $\gamma_2$ and $\mu_2$ for the second one. Therefore, one gets $p = 5$ for computing theoretical chi-squared quantile with significance level $\alpha$ and $K - p - 1$ degrees of freedom. The number of subintervals $K$ is 19. The quantile value is 22.36. The parameters are tuned manually to find suboptimal local minimum for empirical $\chi^2_{emp}$. The chi-squared goodness of fit test gives $\chi^2_{emp} = 9.89$ for the values $\omega = 0.8$, $\gamma_1 = \gamma_2 = 1.14$, $\mu_1 = 13.1$, $\mu_2 = 140$. This is clearly below the threshold of 22.36, so the Weibull distribution is also suitable for approximation (Fig. 6.). The cumulative number of fragments distribution is defined as $N_W\left(\geq x, \bullet\right) = N \cdot \left(1 - F_W\left(m, \bullet\right)\right)$, where dot denotes the list of appropriate arguments.

Next we use the bimodal Grady distribution:

$$F_{GK}\left(m, M_1, \mu_1, M_2, \mu_2\right) = \frac{M_1/\mu_1}{M_1/\mu_1 + M_2/\mu_2}\left[1 - \exp\left(-\frac{m}{\mu_1}\right)\right] + \frac{M_2/\mu_2}{M_1/\mu_1 + M_2/\mu_2}\left[1 - \exp\left(-\frac{m}{\mu_2}\right)\right],$$

where $M_\bullet$ and $\mu_\bullet$ is the subtotal mass and the average mass for the first and second modes respectively. The cumulative number of fragments is described as:

$$N_{GK}\left(m, M_1, \mu_1, M_2, \mu_2\right) = \frac{M_1}{\mu_1}\exp\left(-\frac{m}{\mu_1}\right) + \frac{M_2}{\mu_2}\exp\left(-\frac{m}{\mu_2}\right).$$

Chi-squared test yields the value $\chi^2_{emp} = 15.11$ for the following arguments: $M_1 = 2047.59$, $\mu_1 = 12$, $M_2 = 9237.09$, $\mu_2 = 140$. The threshold for $K = 20$ and $p = 4$ is $\left(\chi^2\right)^{-1}\left(\alpha, K - p - 1\right) = 25$, so the null-hypothesis about Grady distribution is also acceptable with significance level $\alpha$.

The bimodal version of Gilvarry CDF is:

$$F_G(m) = \frac{M_1/\mu_1}{M_1/\mu_1 + M_2/\mu_2}\int_{m_0}^{m}\frac{1}{x}\exp\left(-\frac{x}{\mu_1}\right)dx + \frac{M_2/\mu_2}{M_1/\mu_1 + M_2/\mu_2}\int_{m_0}^{m}\frac{1}{x}\exp\left(-\frac{x}{\mu_2}\right)dx \quad \text{and the corresponding}$$

function for cumulative number of fragments is:

$$N_G(m) = \frac{M_1}{\mu_1}\int_{\max(m, m_0)}^{\infty}\frac{1}{x}\exp\left(-\frac{x}{\mu_1}\right)dx + \frac{M_2}{\mu_2}\int_{\max(m, m_0)}^{\infty}\frac{1}{x}\exp\left(-\frac{x}{\mu_2}\right)dx.$$

The Gilvarry distribution has one special aspect in comparison of other considered distributions. Silvestrov (2004a) emphasize that Gilvarry theory overestimates the number of small lightweight fragments. One can see this on the Fig.8. Goodness-of-fit test yields the value $\chi^2_{emp} = 89.85$ for the $M_1 = 6743.28$, $\mu_1 = 150$, $M_2 = 4541.4$, $\mu_2 = 2270.7$. This is beyond the threshold of $\left(\chi^2\right)^{-1}(\alpha, K-p-1) = 26.3$, with $K = 21$, $p = 4$. Thus, the analyzed data set cannot be approximated by this distribution.

Both Grady and Gilvarry distributions are correct under assumption of nearly-instant singular breaking (Silvestrov, 2004ab). If material is exposed to multiple successive fragmentation events, then the above- mentioned statistical laws are no longer applicable. In this case we must implement the CDF for the fragmentation theory.

The CDF for sequential fragmentation is (Brown, 1989):

$$F(m, \mu, \gamma) = 1 - \frac{N(\geq m)}{N_0} = 1 - \exp\left(-\frac{1}{\gamma+1}\left(\frac{m}{\mu}\right)^{\gamma+1}\right)$$

The corresponding mass distribution has the form:

$$F(m, \mu, \gamma) = \frac{M(\leq m)}{M_0} = 1 - \frac{M(\geq m)}{M_0} = 1 - \Gamma\left(\frac{\gamma+2}{\gamma+1}, \frac{1}{\gamma+1}\left(\frac{m}{\mu}\right)^{\gamma+1}\right) \Big/ \Gamma\left(\frac{\gamma+2}{\gamma+1}\right)$$

where $\Gamma(\alpha, x) = \int_x^\infty t^{\alpha-1}e^{-t}dt$ is a complementary incomplete gamma function. In this mass CDF the ratio of these two gamma functions is essentially the continuous Poisson distribution with parameter $\frac{1}{\gamma+1}\left(\frac{m}{\mu}\right)^{\gamma+1}$ (Ilienko, 2011), obtained by spreading initial discrete probabilistic measure continuously onto $[0; \infty)$. This relation forms the link between discrete distribution of fragments and continuous distribution of masses. There is a similar expression in the paper of (Elek and Jaramaz, 2009), governing the relative (normalized) cumulative number of fragments for another formulation of Weibull distribution:

$$N(m, \mu, \gamma) = \Gamma\left(1 - \frac{1}{\gamma}, \left(\frac{m}{\mu}\right)^{\gamma}\right) \Big/ \Gamma\left(1 - \frac{1}{\gamma}\right).$$

We use binormal variant of sequential fragmentation CDF:

$$F_{SF}(m, \mu_1, \gamma_1, \mu_2, \gamma_2) = 1 - \left[\omega \exp\left(-\frac{1}{\gamma_1+1}\left(\frac{m}{\mu_1}\right)^{\gamma_1+1}\right) + (1-\omega)\exp\left(-\frac{1}{\gamma_2+1}\left(\frac{m}{\mu_2}\right)^{\gamma_2+1}\right)\right].$$

Corresponding complementary cumulative number of fragments is obtained via relation:

$$N_{SF}(\geq m, \mu_1, \gamma_1, \mu_2, \gamma_2) = N_0\left[1 - F_{SF}(m, \mu_1, \gamma_1, \mu_2, \gamma_2)\right]$$

The goodness of fit test gives $\chi^2_{emp} = 21.42$ for the values $\omega = 0.8$, $\gamma_1 = -0.09$, $\mu_1 = 13.1$, $\gamma_2 = -0.01$, $\mu_2 = 121$. The number of independent parameters is $p = 5$, and the number of subintervals $K$ is 18. The value of theoretical chi-squared quantile with significance level $\alpha$ and $K-p-1$ degrees of freedom is 21.03. This threshold is slightly larger than empirical distribution, so the null-hypothesis is still rejected. However, the plot shows good agreement (see Fig.9).

Thus as summarized in the Table 1, the best for use distributions are bimodal lognormal, Weibull and Grady. Wald–Wolfowitz runs test confirms hypothesis for randomness of residual distributions for lognormal and logistic distributions, and rejects it for other distributions. However the runs test is weak in the sense that it would reject this hypothesis even for very precise approximating function approaching to empirical distribution from one side. Kolmogorov-Smirnov test for complex hypothesis is stronger than the runs test, but if the distributions under comparison differ from normal ones, the appropriate Kolmogorov quantile becomes dependent on distribution parameters and function type. For normal-like sigmoid functions and large sample the dependence is small but still can affect acceptance or rejection of likelihood hypothesis. KS-test shows that both bimodal Gilvarry and sequential distributions fail to confirm goodness of fit. Bimodal Grady statistics is slightly above the quantile so formally also does not comply with obtained distribution. Such contradiction between tests demonstrates that statistical estimations are always tricky for real cases and the further discussion of results is desirable.

**Discussion**

The recovered fragments of Košice meteorite together with the recorded trajectory data reported by Borovička et al. (2013) provide an unprecedented opportunity to develop an experimentally justified approach involving proved statistical laws in fragmentation modeling. A reliable fragmentation model is required to interpret number of similar fireball events and to significantly decrease the number of free parameters in future. The application of distribution laws also gives insight about the completeness of fragments recovery within one meteorite fall. Smaller particles can completely vanish in the atmosphere and lead to underestimation of exact number of fragments and meteoroid's pre-entry mass. Mathematical statistics methods could be therefore also used to estimate total meteorite mass based on partial discoveries as well as provide insights on the ablation mechanisms. In this paper we did not focus on modeling the breakup process and analyzed only recovered meteorite fragments. This can be used to estimate the number of key fragmentation events occurred during the atmospheric descent.

Oddershede et al. (1998) and Vinnikov et al. (2014) conducted similar research analyzing the fit of scaled exponential distribution on the meteorites with large number of recovered fragments. They concluded that the scaling parameter in distribution is largely determined by the initial form of the meteoroid than by the atmospheric trajectory. The authors therefore emphasize that such statistical analysis can be used to empirically estimate the initial meteoroid shape by comparison with relevant data of obtained masses.

Thus the distributions considered in this study provide us additional information about Košice meteorite. Our investigation shows good agreement of the anticipated statistical functions with distribution of recovered fragments, especially when bimodality is assumed. According to bimodal lognormal, bimodal Grady, bimodal sequential and bimodal Weibull fragmentation distributions we can assume that two processes took place. One process with the mean fragments mass about 12 g, and another one with the mean mass around 140 g. It can hint on primary singular precursor breaking into two independent pieces with residual masses of approximately 2 and 9 kg respectively (the latter includes two last anomalously heavy pieces of about 2 kg each) at early stages of the entry ahead of later shower fragmentation in lower layers of the atmosphere. We attribute heavy fragments to the second piece since smaller one has insufficient remaining mass to produce two heavy fragments along with corresponding amount of smaller fragments. Thus the lesser ancestor yields about 2 kg of multiple lightweight fragments with the mean around 12 g. The bigger ancestor produces 9 kg of fragments with the mean around 140 g.

The above mentioned sample gap spans from 318 to 2167.4 g. Assuming mutual independence of meteorite fragment masses the probability for a piece to occur within the gap is $F(2167.4,\bullet) - F(318,\bullet)$. Table 2 provides such probabilities for various distributions with suboptimal free parameters. Each singular event is known as Bernoulli trial, and the stochastic sequence of successes/failures is known as binomial distribution. The binomial CDF gives the total probability for the $n$ trials with number of successes no greater than $k$ with singular success probability equal to $p_s$.

The results obtained show that it is very unlikely not to encounter at least one fragment inside the specified gap. The expected frequencies provide the mean number of fragments within the interval. The actual residual masses of anticipated pieces are arbitrary within the bounding values of (318; 2167.4), but from extrapolation of mass ratios and uniform filling of other intervals we can speculate about three missing pieces, e.g. with masses about 500, 800 and 1200g. As an alternative, these pieces can be accounted for by 'dividing' one of the massive fragments. This would support a presence of initial macro-scale cracks within the entering body (Consolmagno and Britt, 1998) or would lead to the conclusion that the large pieces are more durable than small ones and their microstructure differs from all other fragments, what was not confirmed by measurement results reported in Kohout et al. (2014).

The statistical investigation of obtained samples is based on the assumption that even intensive ablation after the fragmentation does not alter resulting cumulative mass distribution. According to Oddershede et al. (1998), ablation becomes negligible after fragmentation took place. We also regard the invariance this of statistical property as highly plausible, at least for chondrite bodies. However it is quite desirable to prove this hypothesis and outline its area of applicability.

It would be useful to conduct further study addressing the correlations between best-fit parameters of statistical distributions for the obtained sample and the experimentally gained corresponding values for the generalized initial shape of the fragmenting brittle body. Dark flight simulations of meteorite fragments can be conducted with respective masses generated via one of the appropriate above mentioned CDFs (Vinnikov et al., 2013).

In this work we assume that the Košice meteorite fragments originate from at least two materials with different structural strength (due to random density of cracks) within one pre-entry body or from two major primary fragmentation events. However, we question the formation of two largest fragments. They are too few to form their own statistically valid distribution and do not fit within the mass distributions of the more lightweight pieces. The most obvious purpose of all presented models is to check if the recovered collection of fragments is complete. Based on our results it is estimated that two to three larger fragments 500-1000g each exist, but may be not recovered or were not reported by illegal meteorite hunters.

**Conclusions**

Based on the recovered meteorite fragments, we have constructed a robust theory for meteoroid mass distribution during fragmentation. Following Borovička et al. (2013) we confirm that the first significant fragmentation of the Košice meteoroid occurred in two phases, which can be described by bimodal lognormal, bimodal Grady, bimodal sequential and bimodal Weibull fragmentation distributions. This conclusion hints either at fragmentation before Earth encounter and atmospheric entry of two independent meteoroids, or at meteoroid collapse in the upper atmosphere due to the presence of initial macro-scale cracks within the entering body. Thus, prior to extensive fragmentation in the lower atmosphere, the Košice meteoroid was represented by two independent pieces: (a) the smaller piece produced in total about 2 kg of multiple lightweight meteorite fragments with the mean around 12 g; (b) the larger piece produced in total about 9 kg of meteorite fragments, including the two heaviest pieces of 2.374 kg and 2.167 kg, with the mean around 140 g.

There are clear indications for the existence of Košice meteorite fragments with the masses ranging from 500 g to 1000 g, which may not have been recovered or, at least, were not officially reported. Our investigation of Košice meteorite fragments suggests the use of the following models as being most appropriate to derive the fragment distributions: Bimodal Weibull, bimodal Grady and bimodal lognormal. Furthermore, the Weibull and Grady functions have a more extensive physical basis and are thus the most highly recommended.

## Acknowledgements

The authors would like to give sincere thanks the reviewers of the earlier version of the manuscript for the number of constructive advices and suggestions. The authors are grateful to Valery N. Tutubalin, Professor of the Department of Probability Theory, Faculty of Mechanics and Mathematics at the Lomonosov Moscow State University, for his preliminary review of the manuscript, fruitful discussion and useful comments he has made. We warmly thank Dr. Ian McCrea, head of the UK EISCAT Support Group at the Rutherford Appleton Laboratory, as well as Prof. A.J. Timothy Jull, Editor in Chief of Meteoritics & Planetary Science, for the preliminary review of the manuscript and language corrections. This work is supported by the Academy of Finland projects No. 260027 and No. 257487, by the Russian Foundation for Basic Research, project No. 13-07-00276, by the Ministry of Education, Youth and Sports of the Czech Republic grant No. LH12079, and by the Slovak Research and Development Agency grant No. APVV-0516-10.

**Figures:**

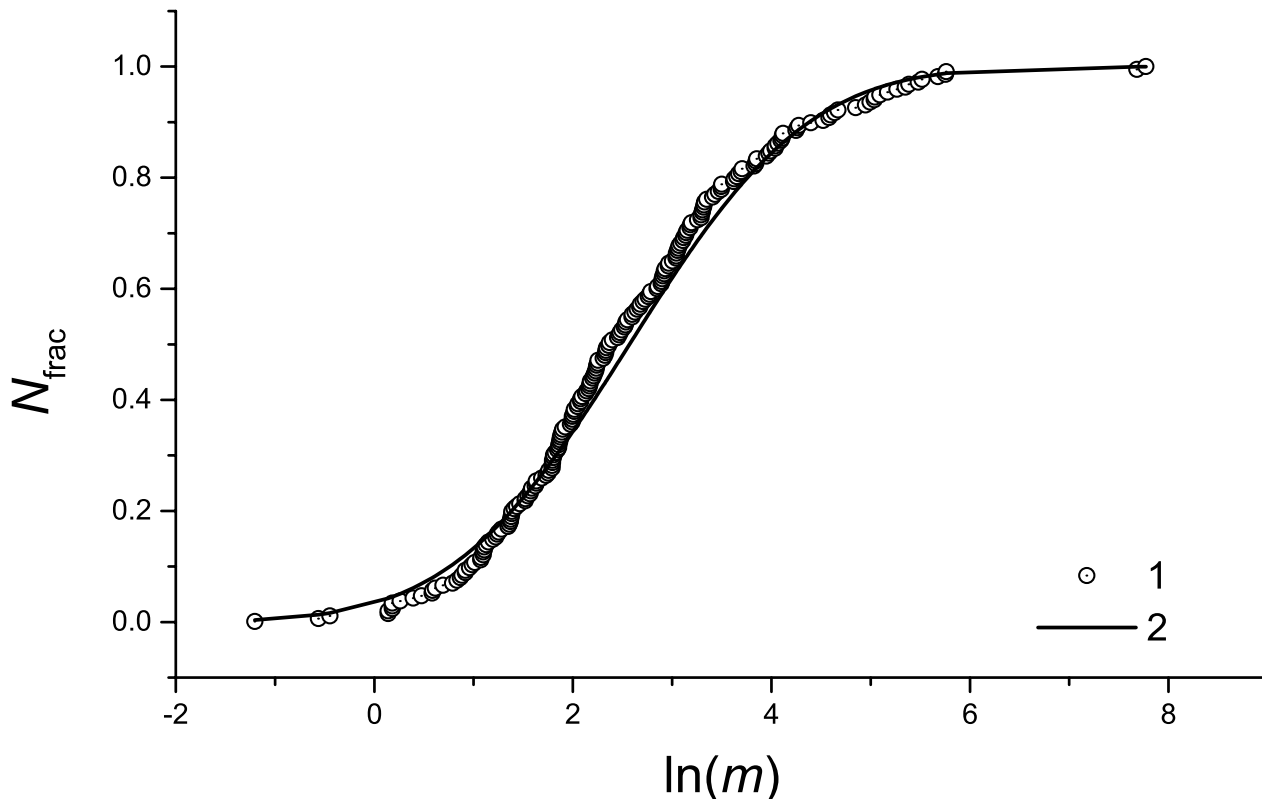

*Fig. 1. The dotes (1) denote observed mass distribution of fragments ($M_{frac}$) less or equal than m (ln(m)); (2) – normal distribution with the mean $\mu = 2.52$ and the standard deviation $\sigma = 1.41$.*

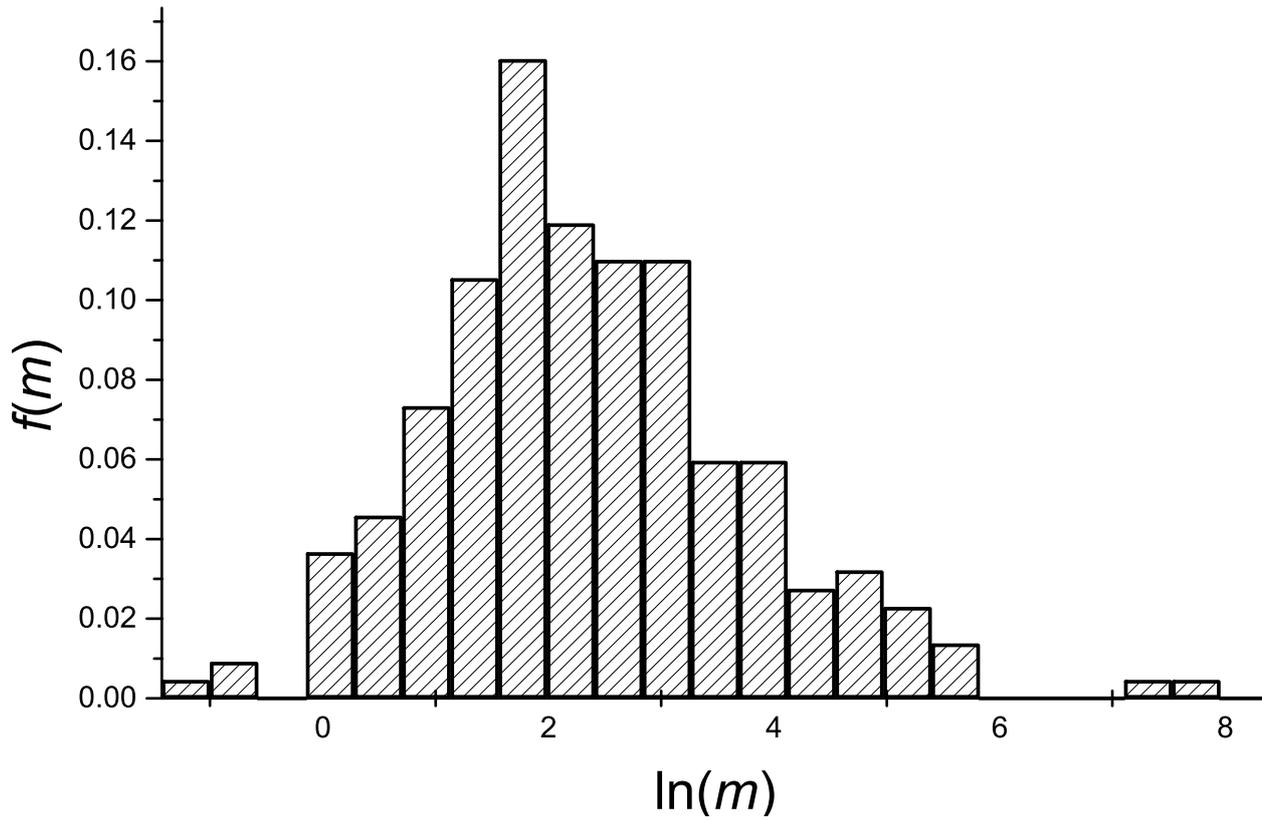

*Fig. 2. Observed mass distribution histogram with 21 uniform sampling mass - ln(m) subintervals vs. fraction of total fragments number f(m).*

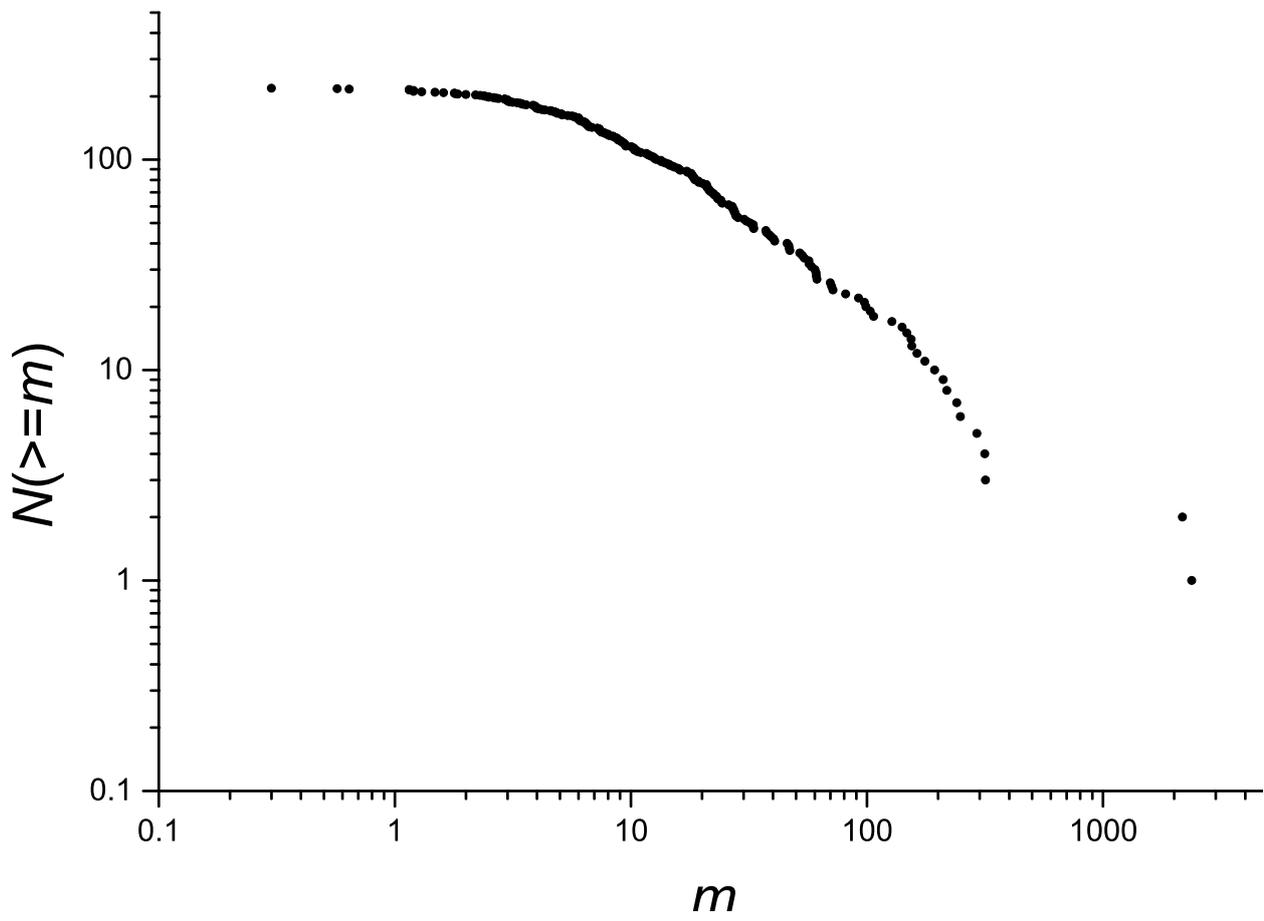

*Fig. 3. Dependence of the complementary cumulative number of fragments $N(\geq m)$ vs m (decimal logarithm scale) for the observed fragment distribution.*

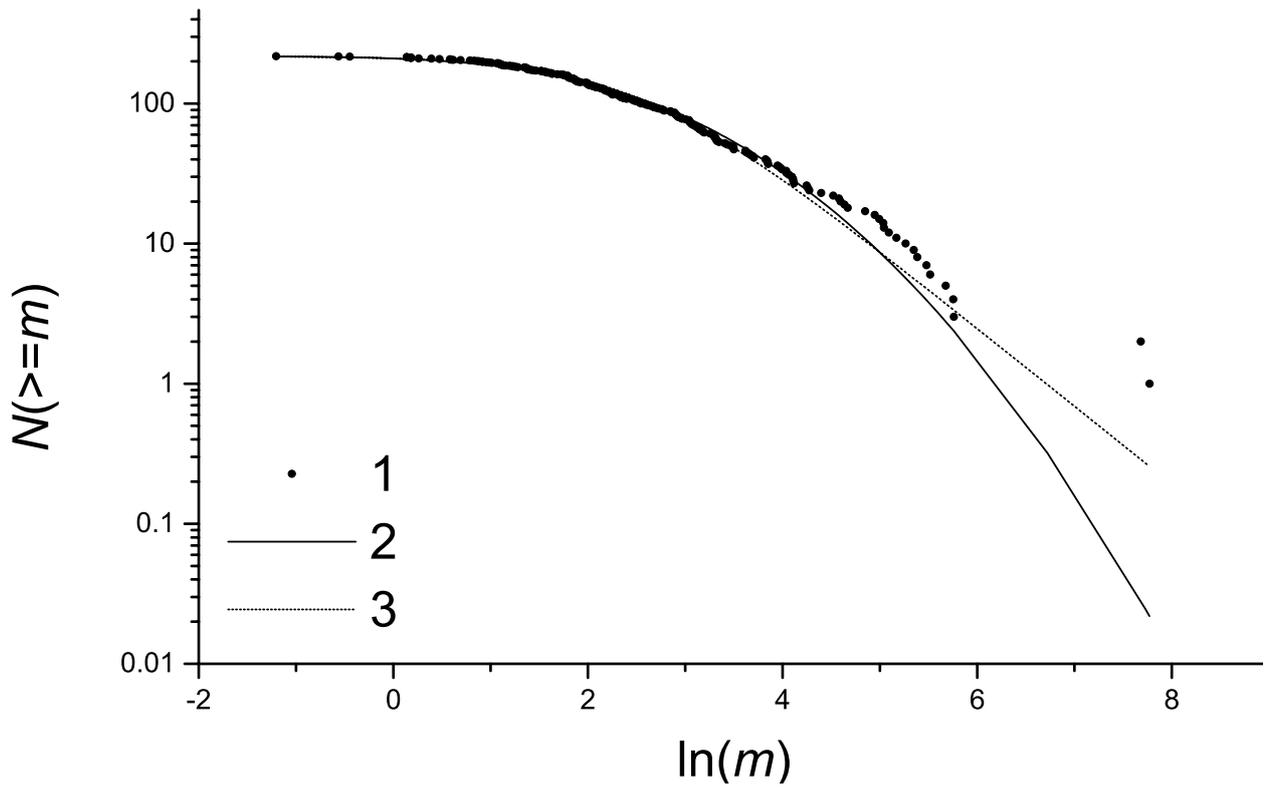

*Fig. 4. Complementary cumulative number of fragments $N(\geq m)$ vs ln(m). 1 – Observed distribution, 2 – Normal distribution for $\ln(m)$ with the mean $\mu = 2.52$ and the standard deviation $\sigma = 1.41$, 3 – Logistic distribution for $\ln(m)$ with the same mean and deviation.*

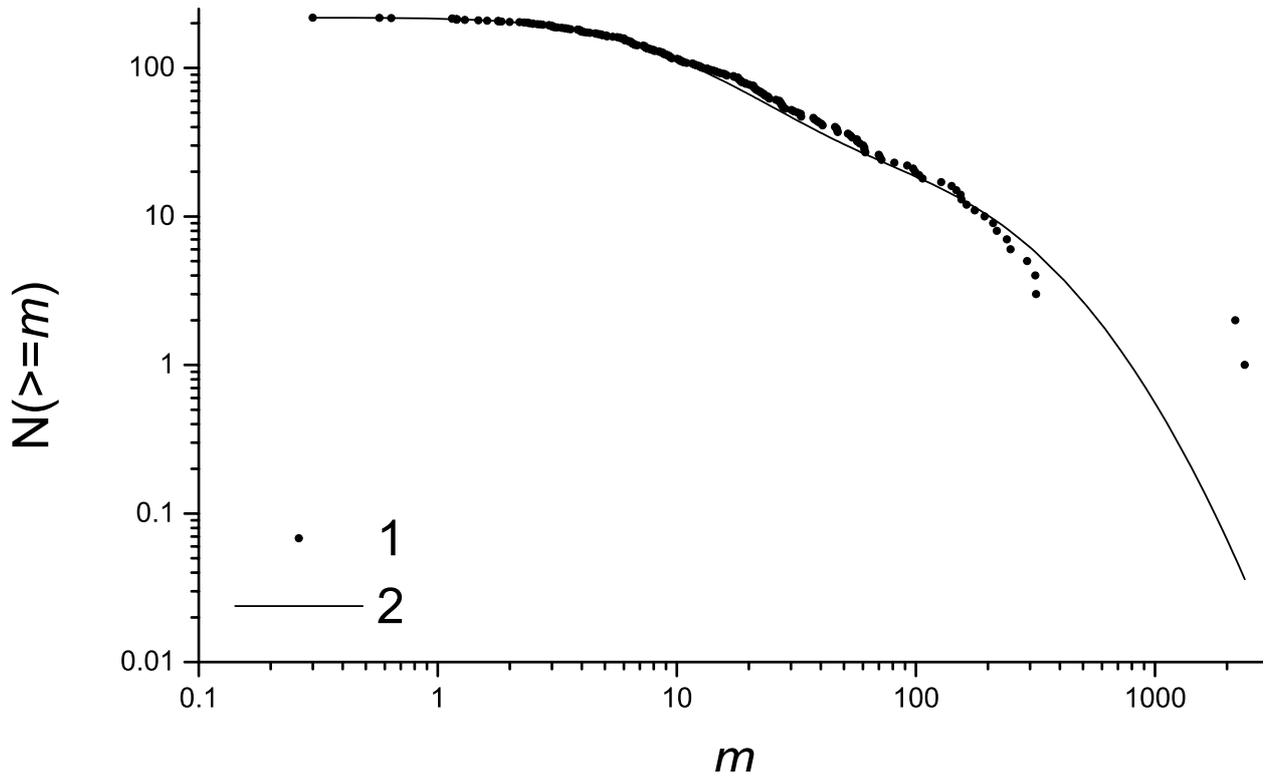

*Fig. 5. Complementary cumulative number of fragments* $N(\geq m)$ *vs m (decimal logarithm scale). 1 – Observed distribution, 2 – Bimodal lognormal distribution with the means* $\mu_1 = 2.22$, $\mu_2 = 5.19$, *the standard deviations* $\sigma_1 = 1.05$, $\sigma_2 = 0.88$ *and* $\omega = 0.9$.

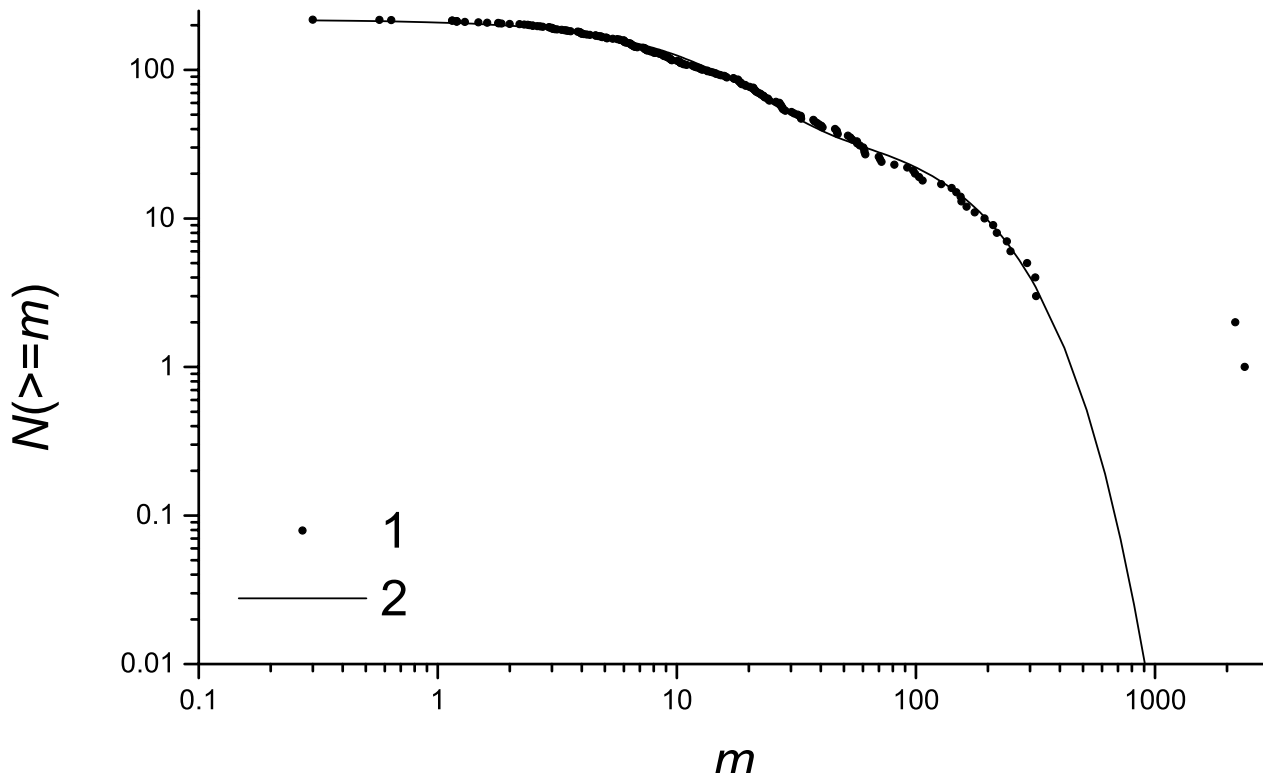

*Fig. 6. Complementary cumulative number of fragments $N(\geq m)$ vs $m$ (decimal logarithm scale) for the sample. (1) – Observed distribution, (2) – Bimodal Weibull distribution with the weighting factor $\omega = 0.8$, $\gamma_1 = \gamma_2 = 1.14$ and $\mu_1 = 13.1$, $\mu_2 = 140$.*

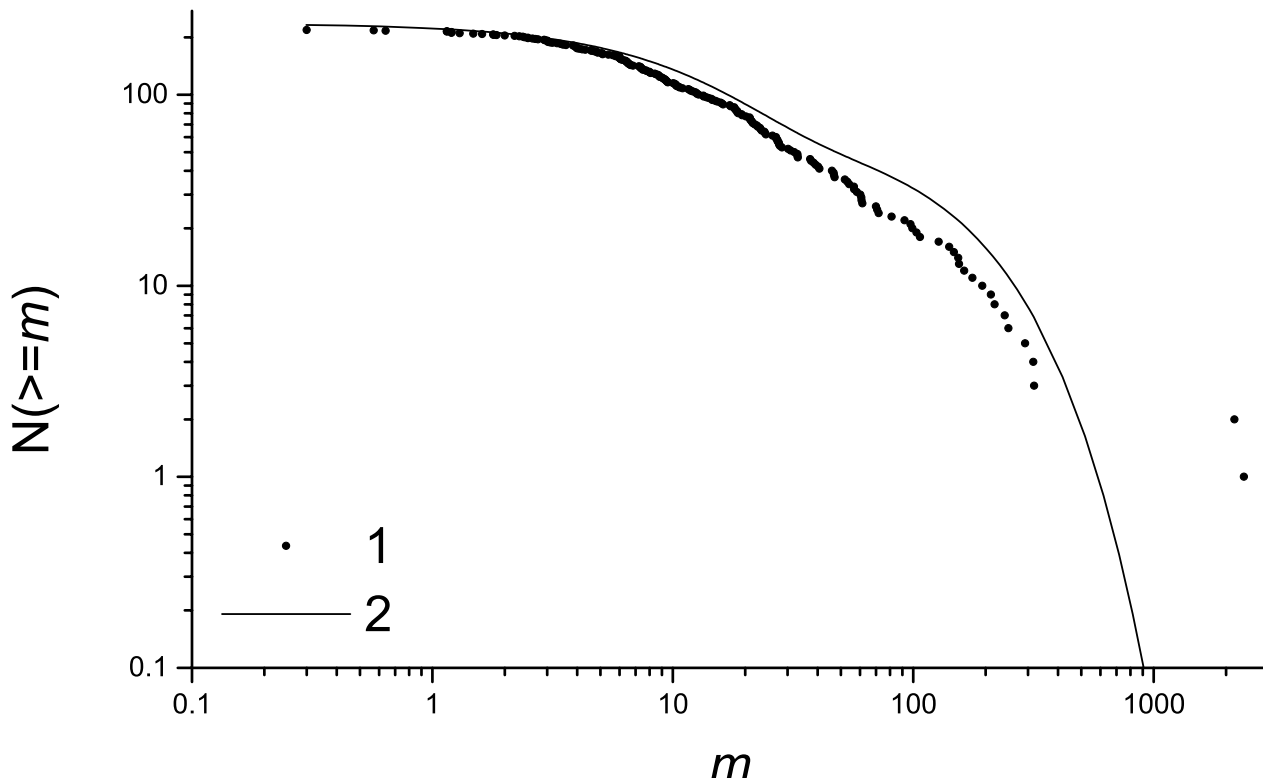

*Fig. 7. Complementary cumulative number of fragments* $N(\geq m)$ *vs m (decimal logarithm scale) for the sample. 1 – Observed distribution, 2 – Bimodal Grady distribution with* $M_1 = 2047.59$, $\mu_1 = 12$, $M_2 = 9237.09$, $\mu_2 = 140$.

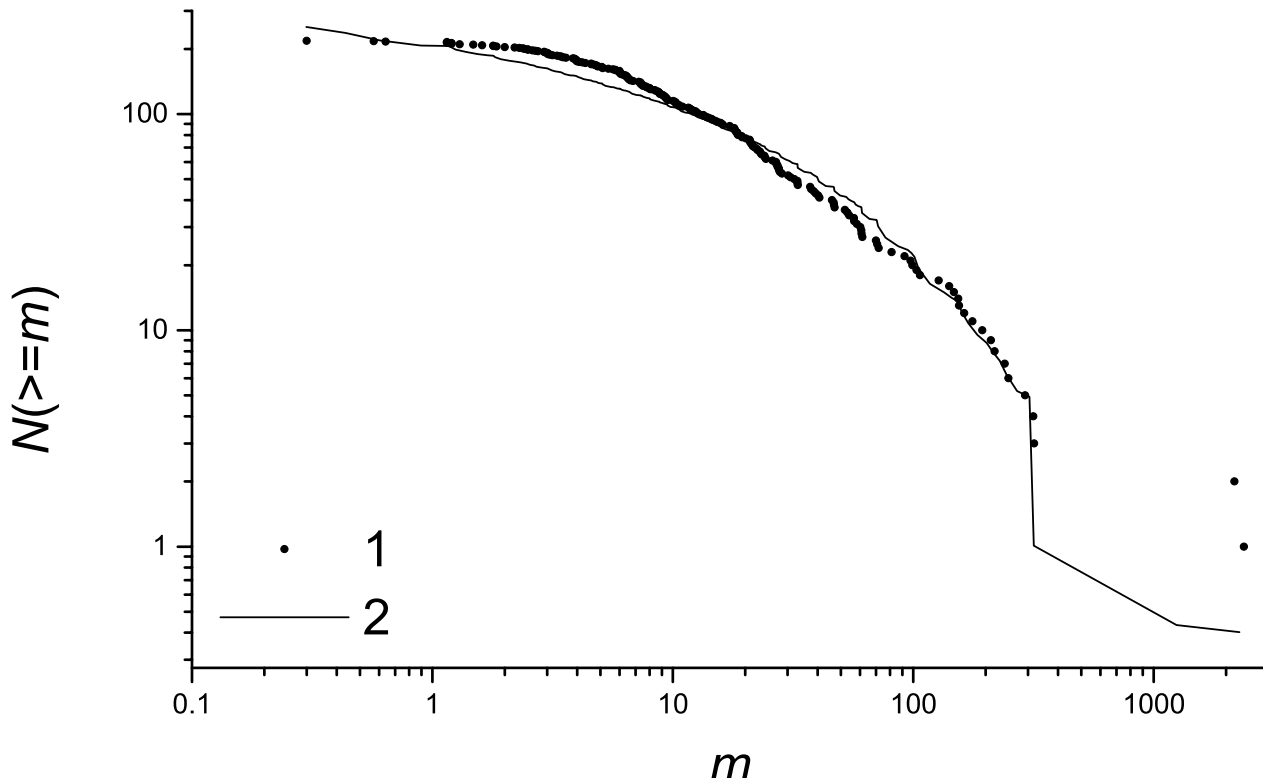

*Fig. 8. Complementary cumulative number of fragments $N(\geq m)$ vs m (decimal logarithm scale) for the sample. 1 – Observed distribution, 2 – Bimodal Gilvarry distribution with $M_1 = 6743.28$, $\mu_1 = 150$, $M_2 = 4541.4$, $\mu_2 = 2270.7$*

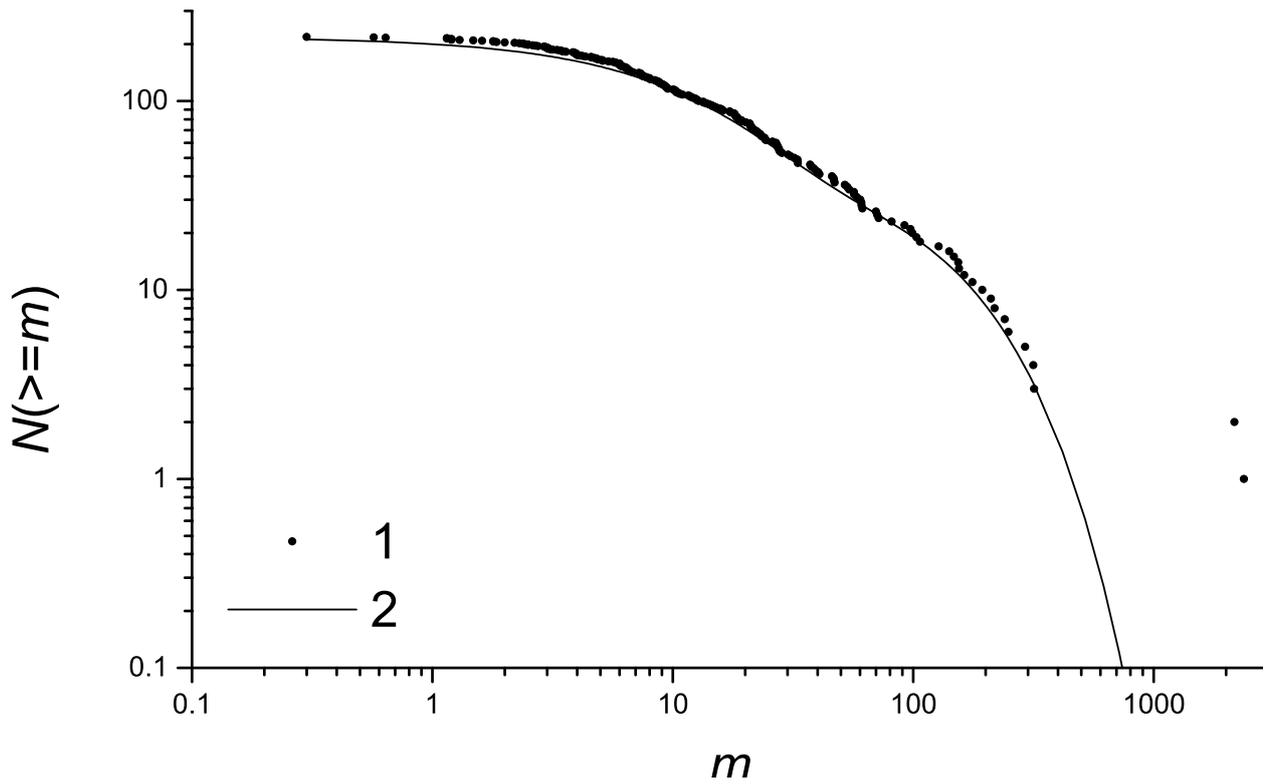

*Fig. 9. Complementary cumulative number of fragments $N(\geq m)$ vs m (decimal logarithm scale) for the sample. 1 – Observed distribution, 2 – Bimodal sequential fragmentation distribution with $\omega = 0.8$, $\gamma_1 = -0.09$, $\mu_1 = 13.1$, $\gamma_2 = -0.01$, $\mu_2 = 121$.*

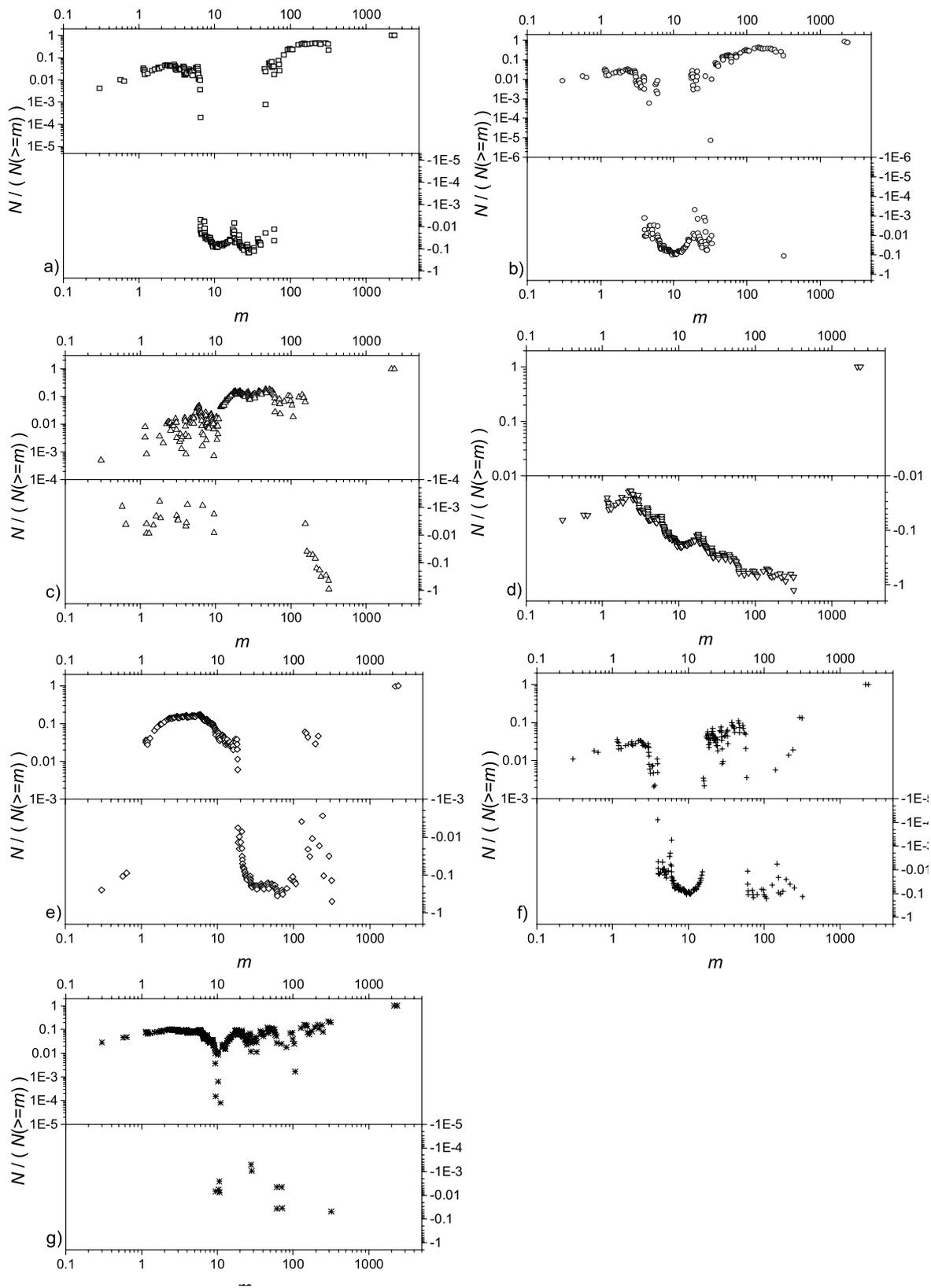

Fig. 10. Relative errors between observed fragment distribution and considered models: a) lognormal, b) logistic, c) bilognormal, d) Grady, e) Gilvarry, f) Weibull, g) sequential fragmentation.

**Tables:**

Table 1. *Pearson's chi-squared test results for the selected distributions.*

| Distribution | Pearson's test value, $\chi^2_{emp}$ | Number of free parameters, p | Number of subintervals, K | Threshold quantile |
|---|---|---|---|---|
| Lognormal | 14.68 | 2 | 19 | 26.3 |
| Logistic | 14.26 | 2 | 19 | 26.3 |
| Lognormal bimodal | 9.44 | 5 | 19 | 22.36 |
| Weibull bimodal | 9.89 | 5 | 19 | 22.36 |
| Grady bimodal | 15.11 | 4 | 20 | 25.0 |
| Gilvarry bimodal | 89.85 | 4 | 21 | 26.3 |
| Sequential bimodal | 21.42 | 5 | 18 | 21.03 |

Table 2. *Probabilities for selected distributions with suboptimal free parameters.*

| Distribution | Probability $p_s$ of singular fragment to occur within the gap (318; 2167.4) | Expected frequency | Probability of complete absence of any fragment from the set within the gap | Probability of 5 or fewer (at least 1) fragments to occur within the gap |
|---|---|---|---|---|
| Lognormal | 0.010740382 | 2.341403304 | 0.094982003 | 0.873568719 |
| Logistic | 0.014009552 | 3.054082381 | 0.046158347 | 0.86580947 |
| Lognormal bimodal | 0.024869714 | 5.421597703 | 0.00412721 | 0.537362366 |
| Weibull bimodal | 0.015649293 | 3.411545776 | 0.032112121 | 0.83856535 |
| Grady bimodal | 0.02876744 | 6.806713155 | 0.001019256 | 0.323890721 |
| Gilvarry bimodal | 0.016412515 | 4.438049085 | 0.011468547 | 0.703942053 |
| Sequential bimodal | 0.014428405 | 3.145392184 | 0.042074882 | 0.860137211 |

# Appendix I

Table 1. Dimensionless masses of known recovered Košice meteorite fragments, *No – number in sequence*

| Mass / g | No | Mass | No | Mass | No | Mass | No |
|---|---|---|---|---|---|---|---|
|  |  | 2.93 | 193 | 4.88 | 166 | 7.36 | 139 |
|  |  | 3 | 192 | 5.06 | 165 | 7.36 | 138 |
| 0.3 | 218 | 3 | 191 | 5.11 | 164 | 7.41 | 137 |
| 0.57 | 217 | 3.03 | 190 | 5.11 | 163 | 7.5 | 136 |
| 0.64 | 216 | 3.03 | 189 | 5.4 | 162 | 7.51 | 135 |
| 1.15 | 215 | 3.09 | 188 | 5.64 | 161 | 7.73 | 134 |
| 1.15 | 214 | 3.16 | 187 | 5.76 | 160 | 7.79 | 133 |
| 1.2 | 213 | 3.3 | 186 | 5.8 | 159 | 8 | 132 |
| 1.2 | 212 | 3.4 | 185 | 6 | 158 | 8 | 131 |
| 1.2 | 211 | 3.45 | 184 | 6 | 157 | 8.1 | 130 |
| 1.3 | 210 | 3.5 | 183 | 6.02 | 156 | 8.4 | 129 |
| 1.48 | 209 | 3.6 | 182 | 6.02 | 155 | 8.5 | 128 |
| 1.61 | 208 | 3.85 | 181 | 6.08 | 154 | 8.67 | 127 |
| 1.79 | 207 | 3.9 | 180 | 6.09 | 153 | 8.74 | 126 |
| 1.8 | 206 | 3.94 | 179 | 6.19 | 152 | 8.8 | 125 |
| 1.85 | 205 | 3.95 | 178 | 6.33 | 151 | 8.83 | 124 |
| 2 | 204 | 3.97 | 177 | 6.4 | 150 | 9.01 | 123 |
| 2.2 | 203 | 4 | 176 | 6.41 | 149 | 9.11 | 122 |
| 2.3 | 202 | 4 | 175 | 6.47 | 148 | 9.2 | 121 |
| 2.38 | 201 | 4.09 | 174 | 6.5 | 147 | 9.3 | 120 |
| 2.42 | 200 | 4.2 | 173 | 6.53 | 146 | 9.37 | 119 |
| 2.5 | 199 | 4.33 | 172 | 6.57 | 145 | 9.41 | 118 |
| 2.5 | 198 | 4.57 | 171 | 6.62 | 144 | 9.5 | 117 |
| 2.61 | 197 | 4.58 | 170 | 6.68 | 143 | 9.53 | 116 |
| 2.68 | 196 | 4.7 | 169 | 6.84 | 142 | 10.04 | 115 |
| 2.75 | 195 | 4.8 | 168 | 7.22 | 141 | 10.2 | 114 |
| 2.92 | 194 | 4.84 | 167 | 7.34 | 140 | 10.3 | 113 |
| 10.32 | 112 | 18.56 | 81 | 32 | 50 | 103.25 | 19 |
| 10.37 | 111 | 18.72 | 80 | 32.98 | 49 | 106.75 | 18 |
| 10.6 | 110 | 19.33 | 79 | 33 | 48 | 127.57 | 17 |
| 10.68 | 109 | 19.41 | 78 | 33.14 | 47 | 141 | 16 |
| 11 | 108 | 20.15 | 77 | 37.3 | 46 | 147.52 | 15 |
| 11.61 | 107 | 20.9 | 76 | 37.6 | 45 | 154 | 14 |
| 11.77 | 106 | 20.93 | 75 | 38.5 | 44 | 155 | 13 |
| 11.9 | 105 | 21 | 74 | 39.3 | 43 | 163 | 12 |
| 12.1 | 104 | 21.22 | 73 | 40.23 | 42 | 176.17 | 11 |
| 12.4 | 103 | 21.4 | 72 | 40.71 | 41 | 193.6 | 10 |
| 12.54 | 102 | 21.55 | 71 | 45.89 | 40 | 210.5 | 9 |
| 12.62 | 101 | 21.95 | 70 | 46.67 | 39 | 218 | 8 |
| 12.81 | 100 | 22.39 | 69 | 46.82 | 38 | 240 | 7 |
| 13.41 | 99 | 22.57 | 68 | 47.12 | 37 | 249 | 6 |
| 13.44 | 98 | 23.03 | 67 | 51.96 | 36 | 292 | 5 |
| 13.8 | 97 | 23.17 | 66 | 53.2 | 35 | 315.91 | 4 |
| 14.13 | 96 | 23.41 | 65 | 54.22 | 34 | 318 | 3 |
| 14.5 | 95 | 24.08 | 64 | 56.7 | 33 | 2167.4 | 2 |
| 14.6 | 94 | 24.17 | 63 | 56.8 | 32 | 2374 | 1 |
| 15.01 | 93 | 24.4 | 62 | 58.3 | 31 | | |
| 15.3 | 92 | 25.97 | 61 | 60.21 | 30 | | |
| 15.8 | 91 | 26.89 | 60 | 60.8 | 29 | | |
| 16 | 90 | 26.9 | 59 | 61.03 | 28 | | |
| 16.2 | 89 | 27.25 | 58 | 61.4 | 27 | | |
| 17.25 | 88 | 27.32 | 57 | 70 | 26 | | |
| 17.38 | 87 | 27.6 | 56 | 70.86 | 25 | | |
| 18 | 86 | 27.7 | 55 | 71.85 | 24 | | |
| 18.06 | 85 | 27.89 | 54 | 81.3 | 23 | | |
| 18.21 | 84 | 28.48 | 53 | 92 | 22 | | |
| 18.3 | 83 | 30.2 | 52 | 97.48 | 21 | | |
| 18.54 | 82 | 30.9 | 51 | 99.1 | 20 | | |

Table 2. Number of recovered fragments in H chondrite meteorites with known orbits.

| Meteorite name | country | year | mass found, kg | type | No of pieces | reference |
|---|---|---|---|---|---|---|
| Příbram | Czechoslovakia | 1959 | 5.8 | H5 | 4 | Ceplecha, 1961; Spurný et al., 2003 |
| Lost City | USA | 1970 | 17.2 | H5 | 4 | McCrosky et al. 1971 |
| Benešov* | Czech Republic | 1991 | 0.002 | H5 | 3 | Spurný et al., 2012 |
| Peekskill | USA | 1992 | 12.6 | H6 | 1 | Brown et al., 1994 |
| Morávka | Czech Republic | 2000 | 1.4 | H5 | 6 | Borovička and Kalenda, 2003 |
| Buzzard Coulee | Canada | 2008 | ~ 41 | H4 | 129 | Milley, 2010 |
| Grimsby | Canada | 2009 | 0.22 | H4-6 | 13 | Brown et al., 2011 |
| Košice | Slovakia | 2010 | 11.3 | H5 | 218 | Borovička et al., 2013 |
| Mason Gully | Australia | 2010 | 0.025 | H5 | 1 | Spurný et al., 2011 |

## Appendix II

## Goodness of fit tests

Pearson's chi-squared test is carried out as follows:

1. The initial sample interval $L = [x_0; x_K]$ is divided into K successive subintervals $L_k = [x_{k-1}; x_k]$ so that $L = \bigcup\limits_{k=1}^{K} L_k$. It is expected that every subinterval contains experimental points and every experimental point belongs to one interval only.

2. Number of points is counted within every subinterval $L_k$ as the sample frequency $N_k$, $\sum\limits_{k=1}^{K} N_k = N$.

3. Theoretical frequencies $\nu_k$ are computed from a probability to place occur a random variable $X$ within the subinterval $L_k$ with regard of asymptotic tails. The applicability for theoretical distribution of $X$ to approximate the sample is the idea behind so called null hypothesis.

$$\nu_1 = N \cdot P\left(X \in (L_1 \cup (-\infty; x_0])\right) = N \cdot F(x_1, \bullet),$$

$$\nu_k = N \cdot P(X \in L_k) = N \cdot [F(x_k, \bullet) - F(x_{k-1}, \bullet)], \ k = 2, \ldots, K-1,$$

$$\nu_K = N \cdot P\left(X \in (L_K \cup [x_K; \infty))\right) = N \cdot [1 - F(x_{K-1}, \bullet)].$$

4. The empirical chi-squared statistics is formed from the calculated data via chi-squared distance: $\chi_{emp}^2 = \sum\limits_{k=1}^{K} \frac{(\nu_k - N_k)^2}{\nu_k}$

5. Obtained statistics $\chi_{emp}^2$ is compared against chi-squared quantile with significance level $\alpha$ and $K - p - 1$ degrees of freedom, where $p$ is the number of parameters for the assumed theoretical distribution with $F(x, \bullet)$ as its CDF.

G-test is an more accurate extension of chi-squared test and is calculated as $G_{emp} = 2 \sum\limits_{k=1}^{K} \left( N_k \ln\left( \frac{N_k}{\nu_k} \right) \right)$. This test didn't receive acknowledgement until the development of fast personal computers, due to complexity of manual multiple calculations of the logarithm function.

Kolmogorov's test computes another distance $D$ between empirical and theoretical CDFs:

$D = \sup\limits_{-\infty < x < \infty} \left| F_{emp}(x) - F(x, \bullet) \right|$, where $F_{emp} = \frac{1}{N} \sum\limits_{i=1}^{N} I(x \le x_i)$ and $I_k(x \le x_i) = \begin{cases} 1, x \le x_i \\ 0, x > x_i \end{cases}$. Then obtained statistics $D$ is compared against desired quantile of Kolmogorov's distribution. In practice, the evaluation of the supremum over continuous $x$ is replaced with the equivalent computation of the maximum over discrete $x_i$:

$$D = \max\limits_{1 \le i \le N} \left[ \frac{i}{N} - F(x_i, \bullet); F(x_i, \bullet) - \frac{i-1}{N} \right].$$

One can also use other methods. For example we consider more simple Wald–Wolfowitz runs test applied to residuals of abovementioned complementary distributions of fragments number. This non-parametric statistical test checks a randomness hypothesis for the error vector defined as $e = \{e_i\}_1^N = \{N_{emp}(x_i) - N(x_i, \bullet)\}_1^N$. The algorithm proceeds as follows:

1. Count the numbers $n_+$, $n_-$ of positive and negative signed $e_i$. Count the number of runs $r$ (sequences of the same sign).

2. If $n_+ > 10$ and $n_- > 10$ compute the mean $\mu = \dfrac{2 n_+ n_-}{n_+ + n_-} + 1$ and the standard deviation as

$$\sigma^2 = \frac{2 n_+ n_- \left(2 n_+ n_- - n_+ - n_-\right)}{\left(n_+ + n_-\right)^2 \left(n_+ + n_- - 1\right)}.$$ For the case of $3 \le n_+ \le 10$, $3 \le n_- \le 10$ see (Draper et al, 1998).

3. Proceed with upper $r > \mu$ and lower $r < \mu$ tail tests: $\zeta_L = \dfrac{r - \mu + 0.5}{\sigma}$, $\zeta_U = \dfrac{r - \mu - 0.5}{\sigma}$. Compute p-value of the standard normal distribution: $2\left(1 - F\left(|\xi_\bullet|, 0, 1\right)\right)$. Compare it against significance level $\alpha$. The hypothesis of residual randomness is rejected if obtained p-value is less then $\alpha$, and accepted otherwise.

However, this test is a complementary to chi-squared test, since the latter deals with the distances but not the signs.

## Appendix III.

Table 1. Pearson's chi-squared test for normal distribution.

| k | $x_k$ | $N_k$ | $v_k$ | $\dfrac{(v_k - N_k)^2}{v_k}$ |
|---|---|---|---|---|
| 0 | -1.214 | – | – | – |
| 1 | 0.222385 | 8 | 11.25128 | 0.939522 |
| 2 | 0.740802 | 7 | 11.29922 | 1.635805 |
| 3 | 1.118383 | 15 | 12.31689 | 0.584488 |
| 4 | 1.397432 | 14 | 11.50578 | 0.540699 |
| 5 | 1.631209 | 11 | 11.15427 | 0.002134 |
| 6 | 1.862518 | 15 | 12.24445 | 0.620122 |
| 7 | 2.048975 | 15 | 10.59152 | 1.834934 |
| 8 | 2.246532 | 16 | 11.74108 | 1.544863 |
| 9 | 2.424881 | 10 | 10.88509 | 0.071969 |
| 10 | 2.636484 | 11 | 13.0244 | 0.314655 |
| 11 | 2.816412 | 8 | 10.96292 | 0.800779 |
| 12 | 3.021477 | 12 | 12.13565 | 0.001516 |
| 13 | 3.225763 | 15 | 11.48688 | 1.074444 |
| 14 | 3.480819 | 12 | 13.20342 | 0.109686 |
| 15 | 3.700543 | 8 | 10.16259 | 0.460199 |
| 16 | 4.052569 | 10 | 13.68793 | 0.993639 |
| 17 | 4.459967 | 9 | 11.82676 | 0.675635 |
| 18 | 5.218622 | 12 | 12.40047 | 0.012933 |
| 19 | 7.783332 | 10 | 6.119386 | 2.460896 |
| Total | | 218 | – | 14.67891501 |

Table 2. Wald–Wolfowitz runs test for selected distributions.

| Distribution | $n_+$ | $n_-$ | $r$ | $\mu$ | $\sigma$ | $\zeta$ |
|---|---|---|---|---|---|---|
| Lognormal | 108 | 110 | 7 | 28.25 | 7.37 | 0.0049 |
| Logistic | 106 | 112 | 17 | 28.23 | 7.36 | 0.1449 |
| Lognormal bimodal | 29 | 189 | 17 | 13.57 | 3.38 | 0.3861 |
| Weibull bimodal | 108 | 110 | 13 | 28.25 | 7.37 | 0.0452 |
| Grady bimodal | 2 | 216 | 2 | N/A | N/A | N/A |
| Gilvarry bimodal | 141 | 77 | 8 | 25.9 | 6.73 | 0.0097 |
| Sequential bimodal | 207 | 11 | 13 | 6.22 | 1.38 | 5.72E-06 |

Table 3. Kolmogorov-Smirnov test for selected distributions. Quantile at level $\alpha = 0.05$ is 0.895.

| Distribution | $\left( \sqrt{N} - 1 + \dfrac{0.85}{\sqrt{N}} \right) D$ |
|---|---|
| Lognormal | 0.664 |
| Logistic | 0.798 |
| Lognormal bimodal | 0.841 |
| Weibull bimodal | 0.861 |
| Grady bimodal | 0.897 |
| Gilvarry bimodal | 13.819 |
| Sequential bimodal | 1.275 |